\documentclass[12pt]{iopart}

\usepackage{graphicx}
\usepackage{graphics}
\usepackage{latexsym}
\usepackage{color}
\usepackage{epsfig}
\usepackage{mathrsfs}
\usepackage{float}
\usepackage{color}
\usepackage{gensymb}
\usepackage{soul} 
\usepackage{stfloats} 
\usepackage{stackrel}

\begin{document}

\title[Backbone diffusion and first-passage dynamics in a comb structure with
confining branches\dots]{Backbone diffusion and first-passage dynamics in a comb structure with
confining branches under stochastic resetting}

\author[R. K. Singh, T. Sandev, A. Iomin and R. Metzler]{R. K. Singh$^{1}$, T. Sandev$^{2,3,4}$, A. Iomin$^{5}$ and R. Metzler$^{2}$}

\address{$^{1}$ Department of Physics, Bar-Ilan University, Ramat-Gan 5290002, Israel \\ $^{2}$ Institute of Physics
\& Astronomy, University of Potsdam, D-14776 Potsdam-Golm, Germany \\ $^{3}$ Research Center for Computer Science and Information Technologies,
Macedonian Academy of Sciences and Arts, Bul. Krste Misirkov 2, 1000 Skopje,
Macedonia \\ $^{4}$ Institute of Physics, Faculty of Natural Sciences
and Mathematics, Ss.~Cyril and Methodius University, Arhimedova 3, 1000
Skopje, Macedonia \\ $^{5}$ Department of Physics, Technion, Haifa 32000, Israel}

\ead{rksinghmp@gmail.com, trifce.sandev@manu.edu.mk, \\ iomin@physics.technion.ac.il, rmetzler@uni-potsdam.de}


\begin{abstract}
We study the diffusive motion of a test particle in a two-dimensional comb
structure consisting of a main backbone channel with continuously distributed
side branches, in the presence of stochastic Markovian resetting
to the initial position of the particle. We assume that the motion along the
infinitely long branches is biased by a confining potential. The crossover
to the steady state is
quantified in terms of a large deviation function, which is derived for the first time for comb structures in present paper. We show that the relaxation region is demarcated by a
nonlinear "light-cone" beyond which the system is evolving in time. We also investigate the first-passage times along the
backbone and calculate the mean first-passage time and optimal resetting rate.
\end{abstract}

%
%
%
%
%

\section{Introduction}

Anomalous is more of a rule rather than an exception. Indeed, the generality of
the statement "anomalous is normal" \cite{katja} is found to hold true time and
again whenever we look at transport in complex and heterogeneous systems. While
most fundamental texts \cite{landau,vankampen} introduce us to normal-diffusive
transport in which the fluctuations grow linearly in time, $\langle x^2(t)\rangle
\simeq t$, as indeed fulfilled for the diffusion of tracer particles in simple
liquids or fragrance molecules in still air, reality teaches us in a very wide
variety of cases \cite{bouchaud,pccp,saxton,hoefling,lenerev} that this linearity
is just a special case of the more general situation of anomalous transport, in which
the mean squared displacement (MSD) takes on the power-law form $\langle x^2(t)
\rangle\simeq t^\alpha$. Here the anomalous diffusion exponent $\alpha$ defines
different diffusive regimes \cite{bouchaud,report}: for $0<\alpha<1$ we talk
about subdiffusion \cite{saxton,jhjeon}, $\alpha=1$ corresponds to normal
diffusion \cite{gillespie}, and the case $\alpha>1$ is referred to as
superdiffusion \cite{caspi,chen,song,shlesinger}. Sometimes for $\alpha>2$ the
term hyperdiffusion is used \cite{haenggi,baskin_iomin, pre102}. We note that the case
$\alpha=1$ in heterogeneous media does not necessarily imply that the process
has a Gaussian probability density function (PDF), instead, for instance,
exponential or stretched Gaussian forms may be observed \cite{granick,erice}.
We also note that the MSD may also grow exponentially, for a multiplicative
noise such as geometric Brownian motion or heterogeneous diffusion processes
\cite{pre2020turbulent,entropy2020,mathematics2021,andreypccp}, or logarithmically in strongly disordered environments \cite{sinai diff1,sinai diff2}. 

A by-now classical model for heterogeneous systems, popularised by Mandelbrot,
are fractals, such as the Sierpi{\'n}ski gasket \cite{mandelbrot,avraham,feder}.
However, such ideal mathematical fractals are often insufficient to adequately
describe real fractals such as networks of rivers, blood vessels, or nerve
fibres, for which random fractals such as percolation clusters are more
appropriate \cite{mandelbrot,avraham,feder}. In many cases such structures have
a characteristic backbone from which various branches emerge \cite{avraham,feder}. A highly effective model addressing transport on such random loopless structures is a comb, in which infinite branches branch off the central backbone. The comb model was introduced to understand anomalous transport in percolation clusters \cite{Ziman,WhBa84,GeGo85,havlin,baskin}. Now, comb-like models are widely employed to describe various experimental applications. Comb-like structures are particularly important
from a biophysical point of view as they provide a way to address transport
along spiny dendrites \cite{santamaria,fedotov,mendez}, in which the transport
properties crucially depend on the underlying geometry \cite{biess}.
Similar approaches are being used in the modelling of river basins with their
often very ramified geometry \cite{calaiori,andrea}. In fact, long time retention
data of tracers in water catchments reveal scaling exponents consistent with
comb dynamics \cite{kirchner,harvey}.

Depending on the specific setting the geometry of comb structures effects both
subdiffusion \cite{iomin2011,lenzijstat,jstat}, including ultraslow diffusion
\cite{mmnp}, and superdiffusion \cite{baskin_iomin,iomin2012,njplenzi}. The
nontrivial nature of transport along a comb is discernible from the fact that
motion along the branches results in a
long-range memory for motion along the backbone which is generically
responsible for the anomalous behaviour of transport \cite{horsthemke}. In
fact, the comb model can be regarded as the discrete version of a continuous
time random walk, in which the return time distribution from a side branch to
the main backbone effects power-law waiting times with diverging mean \cite{havlin},
and thus weak ergodicity breaking and ageing effects \cite{yhe,johannes}. Given the
wide relevance and interesting properties of comb-like, loopless structures, these
represent very powerful mathematical constructs to address motion in heterogeneous
media. We here combine the analysis of diffusion on a comb with the idea of
stochastic resetting.

The concept of stochastic resetting (SR) has attracted considerable attention in
non-equilibrium statistical physics \cite{evans}. In SR a moving particle is reset,
i.e., returned to its initial location at regular or stochastic intervals. This
results in a non-equilibrium steady-state even in cases in which the system under
consideration does not relax to a steady-state in absence of any resets, see, e.g.,
free Brownian motion in $d$ dimensions \cite{snm2014,eule}. The effect of resetting
is particularly relevant for the first-passage properties of the motion of interest
\cite{redner,metzler,bray}. Indeed, even in generic cases in finite domains the
probability density of first-passage times is remarkably broad, and the typical
first-passage time often orders of magnitude smaller than the mean first-passage
time, the latter being sampled by relatively extreme events \cite{aljaz,denis}.
In nature, on the scale of molecular regulation in biological cells this defocusing
is prevented by designed short distances between interacting genes \cite{kolesov,otto}
or by cutting off long first-passage times via inactivation of the respective
regulatory molecules \cite{inact_plos}. Another example comes from the search of
larger animals for food, in which resetting to locations of previous search success
is a typical element of the search process, see \cite{animalreset} and references
therein. Indeed, SR is a powerful way to reduce the first-passage times \cite{evans,pal_search}.
In this sense SR
can "tame the violent" fluctuations in first-passage times thereby reducing the
mean time to reach a threshold in a nontrivial manner \cite{pal2017}. Notably SR
leads to universal fluctuations of first-passage times \cite{reuveni}.

SR can be phrased as a renewal \cite{chechkin,anna} or a non-renewal \cite{anna1}
process. Moreover, SR dynamics was studied for motion in bounded domains
\cite{christou,prasad}, as well as in monotonic \cite{mathematics2021,ray1,ray} and non-monotonic
potentials \cite{pal2015,rk2020}, and under time-dependent resetting \cite{evans2015}. Efficient escape under resetting was investigated
\cite{snm2011}, and it was shown that interesting phase transitions occur in the
parameter space for the optimal resetting rate \cite{animalreset,christou,kusmierz,
campos,das}. A dynamical phase transition was revealed in relaxation to the non-equilibrium steady-state \cite{schehr}. Recently the concept of resetting by random amplitudes has been put forward \cite{metzler arxiv}. Further aspects
revealed in SR are collected in a recent review \cite{jphysa2020}.

The dynamics effected by combining the comb model for the description of
diffusion in loopless heterogeneous structures with SR was studied recently
\cite{prr}, unveiling various transport properties. The 3$d$ comb
considered in \cite{prr} was based on branches of infinite length. However,
in any real system the branches are expected to have a finite size, and the
mean time $\langle t\rangle$ a particle spends in a branch is finite. Of
course, when the resetting rate $r$ in a combined process is high, such
that $\langle t\rangle\gg1/r$, the finite size of the branches can be
neglected. Here we study the case of a general resetting rate in infinite
branches, in which a potential directed towards the backbone ensures finite
mean residence times in the branches. We set up the general equation of
motion for a two-dimensional comb with a backbone along the $x$-axis and
branches extending orthogonally in $y$-direction in section~\ref{sec2} and
study the crossover to the non-equilibrium steady state in section~\ref{sec3}. The
process is characterised in terms of the particle PDF and the MSD. Concretely,
we find that the transport along the backbone crosses over from short-term
anomalous diffusion along the backbone, due to the residence of the particle
in the branches, to normal diffusion at time scales beyond the mean residence
time in the branches. We show that resetting further tames the spread along
the backbone, and the system is found to eventually relax to a steady-state
governed by the geometry and resetting rates. We also address the relaxation
to the steady-state by analysing the associated dynamical phase transition. In
section~\ref{sec4} we then consider the first-passage dynamics in terms of
the first-passage time density, the mean first-passage time and the statistic of zero-crossings which depend
on higher order correlations \cite{lim}. This provides a mean to study the
effects of confinement along the branches and see how resetting affects
the underlying escapes.  We draw our conclusions in section~\ref{sec5}. In
\ref{app1} we develop an alternative viewpoint in terms of a coupled
Langevin equation approach with subordination, while in \ref{app2} we give additional explanation of the confinement along the branches of the comb.

\section{Resetting dynamics in a potential}
\label{sec2}

We consider a comb structure, whose backbone is described by the $x$-axis and
whose branches extend along the $y$-axis. For a test particle performing Brownian
motion in this two-dimensional comb structure with a potential $V=V(y)$ along the
branches, the Fokker-Planck equation describing the dynamics of the PDF $p_r(x,y,t)$
under a constant resetting rate $r$ reads 
(see Refs.~\cite{prr,lenzi} for the case without potential)
\begin{eqnarray}
\label{fpe}
\frac{\partial}{\partial t}p_r(x,y,t)&=&D_x\delta(y)\frac{\partial^2}{\partial x^2}
p_r(x,y,t)+\left(\frac{\partial}{\partial y}V'(y)+D_y\frac{\partial^2}{\partial y^2}
\right)p_r(x,y,t)\nonumber\\&&-rp_r(x,y,t)+r\delta(x-x_0)\delta(y),
\end{eqnarray}
where $p_r(x,y,t=0)=\delta(x-x_0)\delta(y)$, and we choose the potential function to
be piecewise linear,
\begin{equation}
\label{vx}
V(y)=\left\{\begin{array}{ll}- U_0y, & y\le0, \\[0.2cm] U_0y, & y\ge0.\end{array}\right.
\end{equation}
Here we assume that the particle is reset to its initial position $(x_0,0)$ at
a constant resetting rate $r$. Each resetting event to the initial position $x_0$
renews the process at a rate $r$, i.e., between two consecutive renewal events the
particle undergoes diffusion on the comb in the non-monotonic potential (\ref{vx})
along the branches. The last two terms on the right-hand side of equation (\ref{fpe})
represent the loss of probability from the position $(x,y)$ due to the reset to the
initial position $(x_0,0)$, and the probability gain at $(x_0,0)$ due to resetting
from all other positions, respectively. The term $\delta(y)$ implies that the
diffusion along the $x$-direction is allowed only at $y=0$ (the backbone). In this
sense the branches have the role of traps, as explained in the original paper by
Weiss and Havlin \cite{havlin}.

\begin{figure}
\includegraphics[width=8cm]{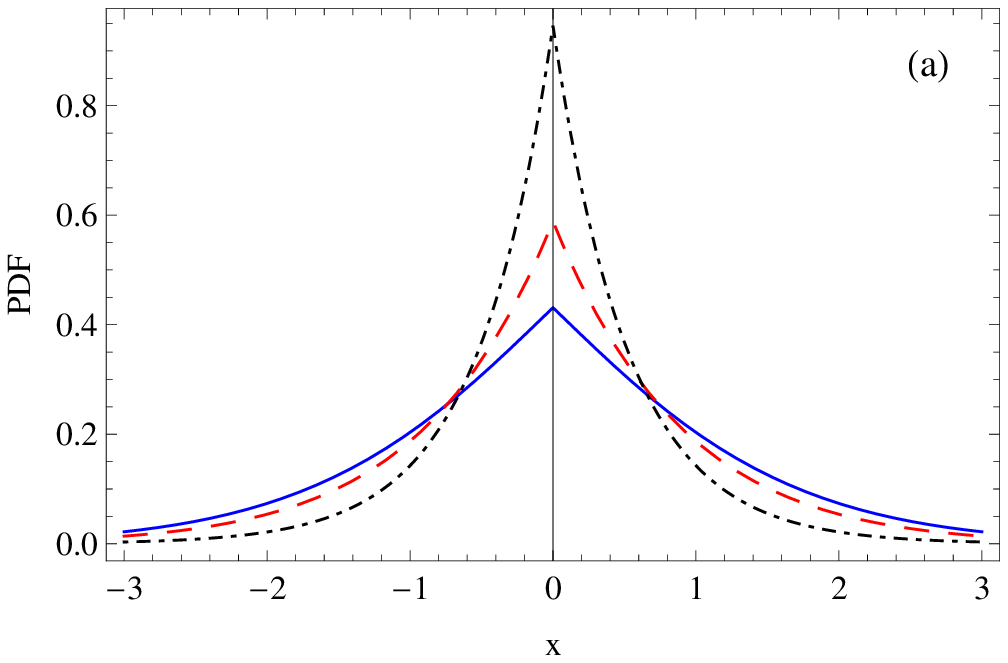}\includegraphics[width=8cm]{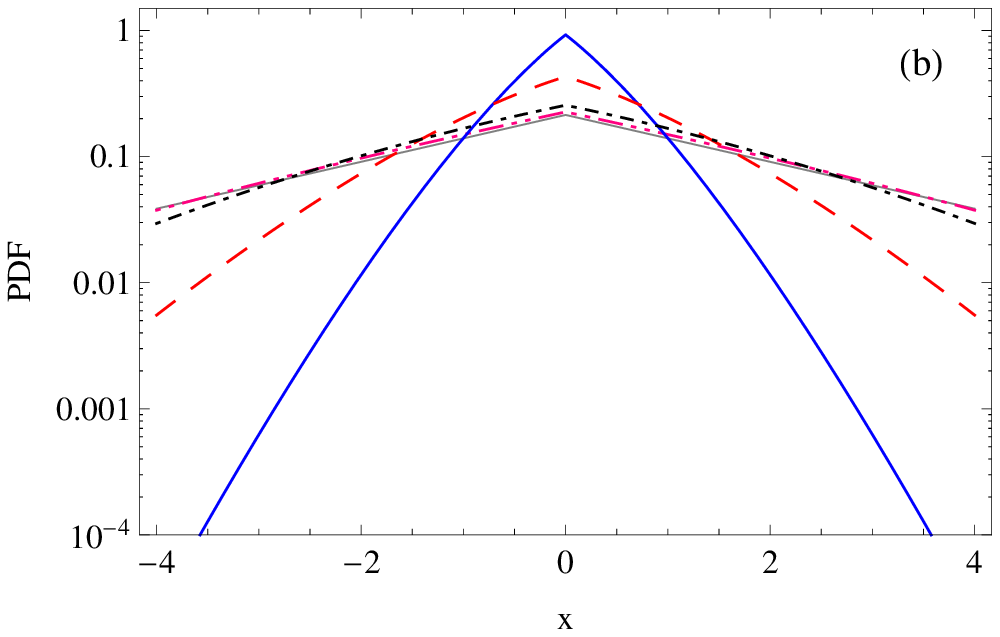}
\caption{PDF (\ref{p1 final laplace U0}) as function of $x$ for (a) time $t=1$ and
resetting rates $r=0.1$ (blue solid line), $r=1$ (red dashed line), $r=5$ (black
dot-dashed line); (b) $r=0.1$ and $t=0.1$ (blue solid line), $t=1$ (red dashed
line), $t=5$ (black dot-dashed line), $t=10$ (violet dot-dot-dashed line), which
approaches the stationary distribution (\ref{p1 statinary}) (solid thin grey line).
We set $D_x=1$, $D_{y}=1$ and $U_0=1$.}
\label{fig1}
\end{figure}

Applying a Laplace transform, $\mathscr{L}\left\{f(t)\right\}=\int_0^{\infty}f(t)
e^{-st}dt=\tilde{f}(s)$ with respect to time $t$ to the dynamic equation (\ref{fpe})
we obtain
\begin{eqnarray}
\fl s\tilde{p}_r(x,y,s)-\delta(x-x_0)\delta(y)&=&D_x\delta(y)\frac{\partial^2}{\partial
x^2}\tilde{p}_r(x,y,s)\nonumber\\&&+\left(U_0\,\mathrm{sign}(y)\frac{\partial}{\partial y}+2U_0
\delta(y)+D_y\frac{\partial^2}{\partial y^2}\right)\tilde{p}_r(x,y,s)\nonumber\\&&-r\tilde{p}_r(x,y,s)+\frac{r}{s}\delta(x-x_0)\delta(y),
\label{pxs}
\end{eqnarray}
with $p_r(x,y,t=0) = \delta(x-x_0)\delta(y)$. Reflecting the symmetry of the system,
equation (\ref{pxs}) is symmetric with respect to $y$-inversion, $y\to-y$. Thus,
after substitution $z=|y|$ we find the following system of
equations
\begin{eqnarray}
&&(s+r)\tilde{p}_r(x,z,s)=U_0\frac{\partial}{\partial z}\tilde{p}_r(x,z,s)+D_y\frac{
\partial^2}{\partial z^2}\tilde{p}_r(x,z,s),\label{syst10}\\
&&-s^{-1}(s+r)\delta(x-x_0)=\left(D_x\frac{\partial^2}{\partial x^2}+2U_0+2D_y\frac{
\partial}{\partial z}\right)\left.\tilde{p}_r(x,z,s)\right|_{z=0}.
\label{syst1}
\end{eqnarray}
From equations (\ref{syst10}) and (\ref{syst1}) we find the solution in the form
\begin{equation}
\label{p g}
\tilde{p}_r(x,y,s)=\tilde{g}_r(x,s)\times\exp\left(-\frac{U_0}{2D_y}\left[1+\Delta
_{s+r}\right]z\right),
\end{equation}
where $\Delta_{s+r}=\sqrt{1+4D_y(s+r)/U_0^2}$. Therefore, for the marginal PDF along
the backbone we have
\begin{equation}
\label{p1 g}
\tilde{p}_{r,1}(x,s)=\int_{-\infty}^{\infty}\tilde{p}_r(x,y,s)dy=2\int_0^{\infty}
\tilde{p}_r(x,z,s)dz=\frac{4D_y}{U_0}\frac{\tilde{g}_r(x,s)}{1+\Delta_{s+r}}.
\end{equation}
From equations (\ref{syst1}), (\ref{p g}), and (\ref{p1 g}) we then obtain
\begin{equation}
\label{p1 eq laplace}
s\tilde{p}_{r,1}(x,s)-\delta(x-x_0)=\frac{D_x}{4D_y}U_0s\times\frac{1+\Delta_{s+r}}
{s+r}\frac{\partial^2}{\partial x^2}\tilde{p}_{r,1}(x,s),
\end{equation}
and by the inverse Laplace transform we obtain the generalised diffusion equation
\cite{csf}
\begin{equation}
\label{p1 eq}
\frac{\partial}{\partial t}{p}_{r,1}(x,t)=\frac{D_x}{2\sqrt{D_y}}\frac{\partial}
{\partial t}\int_0^t\eta(t-t')\frac{\partial^2}{\partial x^2}p_{r,1}(x,t')dt'
\end{equation}
with the memory kernel $\eta(t)$, which is determined by the inverse Laplace
transform\footnote{An alternative consideration based on a subordination approach
is presented in \ref{app1}.} 
\begin{equation}
\label{eta comb potential}
\tilde{\eta}(s)=\frac{1}{s+r}\left[\frac{U_0}{2\sqrt{D_y}}+\left(s+r+\frac{U_0^2}
4D_y\right)^{1/2}\right].
\end{equation}
In time domain this memory kernel reads
\begin{equation}
\eta(t)=\frac{U_0}{2\sqrt{D_y}}e^{-r\,t}+e^{-r\,t}\left(\frac{\exp\left(-\frac{U_0^2}
{4D_y}t\right)}{\sqrt{\pi t}}+\frac{U_0}{2\sqrt{D_y}}\mathrm{erf}\left(\frac{U_0}
{2\sqrt{D_y}}\sqrt{t}\right)\right).
\end{equation}
Fourier-Laplace transforming equation (\ref{p1 eq}) we find
\begin{equation}
\label{eq general fl}
\tilde{p}_{r,1}(k,s)=\frac{\frac{1}{s\eta(s)}}{\frac{1}{\eta(s)}+\frac{D_x}{2\sqrt{
D_y}}k^2},
\end{equation}
which by inverse Fourier transform yields
\begin{equation}
\tilde{p}_{r,1}(x,s)=\frac{1}{2s}\sqrt{\frac{2\sqrt{D_y}}{D_x\tilde{\eta}(s)}}\times
\exp\left(-\sqrt{\frac{2\sqrt{D_y}}{D_x\tilde{\eta}(s)}}|x-x_0|\right).
\end{equation}
More explicitly, after substituting for the memory kernel,
\begin{eqnarray}
\label{p1 final laplace U0}
\tilde{p}_{r,1}(x,s)&=&\frac{1}{2}\sqrt{\frac{2\sqrt{D_y}}{D_x}}\frac{s^{-1}(s+r)^{1/2}}
{\sqrt{\left(s+r+\frac{U_0^2}{4D_y}\right)^{1/2}+\frac{U_0}{2\sqrt{D_y}}}}\nonumber\\&&\times
\exp\left(-\sqrt{\frac{2\sqrt{D_y}}{D_x}}\frac{(s+r)^{1/2}|x-x_0|}{\sqrt{\left(s+r+
\frac{U_0^2}{4D_y}\right)^{1/2}+\frac{U_0}{2\sqrt{D_y}}}}\right).
\end{eqnarray}
According to  the final value theorem of the Laplace transformation, in the long time
limit ($s\rightarrow0$) the stationary distribution reads
\begin{eqnarray}
\nonumber
&&p_{r,1,\mathrm{st}}(x)=\lim_{t\to\infty}p_{r,1}(x,t)=\lim_{s\to0}s\tilde{p}_{r,1}
(x,s)\\
&&=\frac{\frac{r^{1/2}}{2}\sqrt{\frac{2\sqrt{D_y}}{D_x}}}{\sqrt{\left(r+\frac{U_0^2}
{4D_y}\right)^{1/2}+\frac{U_0}{2\sqrt{D_y}}}}\times\exp\left(-\sqrt{\frac{2\sqrt{D_y}
}{D_x}}\frac{r^{1/2}\,|x-x_0|}{\sqrt{\left(r+\frac{U_0^2}{4D_y}\right)^{1/2}+\frac{
U_0}{2\sqrt{D_y}}}}\right).\nonumber\\
\label{p1 statinary}
\end{eqnarray}
For the unconfined case $U_0=0$, we recover from equation (\ref{p1 final laplace U0})
the result for the PDF along the backbone in the case of diffusion in a comb with
stochastic resetting in absence of the potential \cite{prr,lenzi},
\begin{eqnarray}
\label{p1 final laplace}
\tilde{p}_{r,1}(x,s)=\frac{s^{-1}(s+r)^{1/4}}{2}\sqrt{\frac{2\sqrt{D_y}}{D_x}}\times
\exp\left(-\sqrt{\frac{2\sqrt{D_y}}{D_x}}(s+r)^{1/4}|x-x_0|\right),\nonumber\\
\end{eqnarray}
and the corresponding stationary distribution
\begin{equation}
\label{p1 stationary u=0}
p_{r,1,\mathrm{st}}(x)=\frac{1}{2}\sqrt{\frac{2\sqrt{D_y}}{D_x}}r^{1/4}\times\exp
\left(-\sqrt{\frac{2\sqrt{D_y}}{D_x}}r^{1/4}|x-x_0|\right).
\end{equation}
Note that in absence of resetting, the system does not reach stationarity, as can
be seen from equation (\ref{p1 final laplace U0}) by setting $r=0$. In that case
we have
\begin{eqnarray}
\nonumber
\tilde{p}_{0,1}(x,s)&=&\frac{1}{2}\sqrt{\frac{2\sqrt{D_y}}{D_x}}\frac{s^{-1/2}}
{\sqrt{\left(s+\frac{U_0^2}{4D_y}\right)^{1/2}+\frac{U_0}{2\sqrt{D_y}}}}\nonumber\\&&\times 
\exp\left(-\sqrt{\frac{2\sqrt{D_y}}{D_x}}\frac{s^{1/2}|x-x_0|}{\sqrt{\left(s+
\frac{U_0^2}{4D_y}\right)^{1/2}+\frac{U_0}{2\sqrt{D_y}}}}\right)\nonumber\\
&\stackrel[s\to0]{}{\simeq}&\frac{1}{2}\sqrt{\frac{2D_y}{D_xU_0}}s^{-1/2}\times
\exp\left(-\sqrt{\frac{2D_y}{D_xU_0}}s^{1/2}\,|x-x_0|\right),
\label{p1 final laplace U0 r=0}
\end{eqnarray}
from where, by the inverse Laplace transform we obtain the Gaussian PDF \cite{report}
\begin{equation}
\label{gaussian}
p_{0,1}(x,t)=\frac{1}{\sqrt{4\pi D_1t}}\times\exp\left(-\frac{
(x-x_0)^2}{4D_1t}\right),
\end{equation}
where $D_1=\frac{D_xU_0}{2D_y}$ is the diffusion coefficient. It will be shown later that the MSD in the long time limit corresponds to normal
diffusion in absence of resetting. A graphical representation of the PDF and the
transition to the steady state is shown in figure~\ref{fig1}. From figure~\ref{fig1}(a) we observe the cusp at the resetting point $x_0=0$ since the resetting mechanism introduces a source of probability at $x_0=0$. At this point the first derivative is discontinuous. In figure~\ref{fig1}(b) we see that at $t=\frac{1}{r}=10$ the stationary distribution (\ref{p1 statinary}) is almost reached.

From equation (\ref{eq general fl}) we derive the MSD via the relation $\langle x^2
(t)\rangle=\mathscr{L}^{-1}\left.\left\{-\partial^2\tilde{p}_{r,1}(k,s)/\partial
x^2\right\}\right|_{k=0}$ \cite{csf}
\begin{eqnarray}
\nonumber
\langle x^2(t)\rangle&=&2\left(\frac{D_x}{2\sqrt{D_y}}\right)\mathscr{L}^{-1}
\left\{s^{-1}\tilde{\eta}(s)\right\}\\
&=&\frac{D_x}{2D_y}U_0\frac{1-e^{-rt}}{r}-\frac{D_x}{2D_y}U_0\frac{e^{-rt}}{r}
\mathrm{erf}\left(\frac{U_0}{2\sqrt{D_y}}\sqrt{t}\right)\nonumber\\&&+\frac{D_x}{2D_y}U_0
\frac{\Delta_r}{r}\mathrm{erf}\left(\frac{U_0}{2\sqrt{D_y}}\Delta_r\sqrt{t}\right),
\label{msd final}
\end{eqnarray}
where $\Delta_r=\sqrt{1+4D_yr/U_0^2}$.

Let us consider two relevant limiting cases. In absence of confinement, $U_0=0$,
the MSD reads
\begin{equation}
\label{msd final U0=0}
\langle x^2(t)\rangle=\frac{D_x}{\sqrt{D_y}}\frac{\mathrm{erf}\left(\sqrt{rt}
\right)}{\sqrt{r}},
\end{equation}
as it should be for diffusion in a comb with stochastic resetting in absence of
a potential \cite{prr,lenzi}. Conversely, in the absence of resetting ($r=0$) the MSD
(\ref{msd final}) turns to
\begin{eqnarray}
\label{msd final r=0}
\fl\langle x^2(t)\rangle=\frac{D_x}{2D_y}U_0\left[t+\frac{2\sqrt{D_y}}{U_0}
\frac{t^{1/2}}{\Gamma(1/2)}\exp\left(-\frac{U_0^2}{4D_y}t\right)+\left(t
+\frac{2D_y}{U_0^2}\right)\mathrm{erf}\left(\frac{U_0}{2\sqrt{D_y}}\sqrt{t}
\right)\right],\nonumber\\
\end{eqnarray}
which in the long time limit behaves as $\langle x^2(t)\rangle\simeq t$, also confirmed by the Gaussian PDF (\ref{gaussian}). This means
that due to the confining potential along the branches the particle returns back
to the backbone more frequently, resulting in normal diffusion along the $x$-axis.
In this sense the confining potential is an integral part of the resetting
mechanism. It is known that the stochastic resetting of a particle from the branch
to the backbone also leads to normal diffusion along the $x$-axis \cite{prr}. An additional explanation of the confinment along the branches and resulting normal diffusion along the backbone is given in \ref{app2}.

\begin{figure}
\includegraphics[width=8cm]{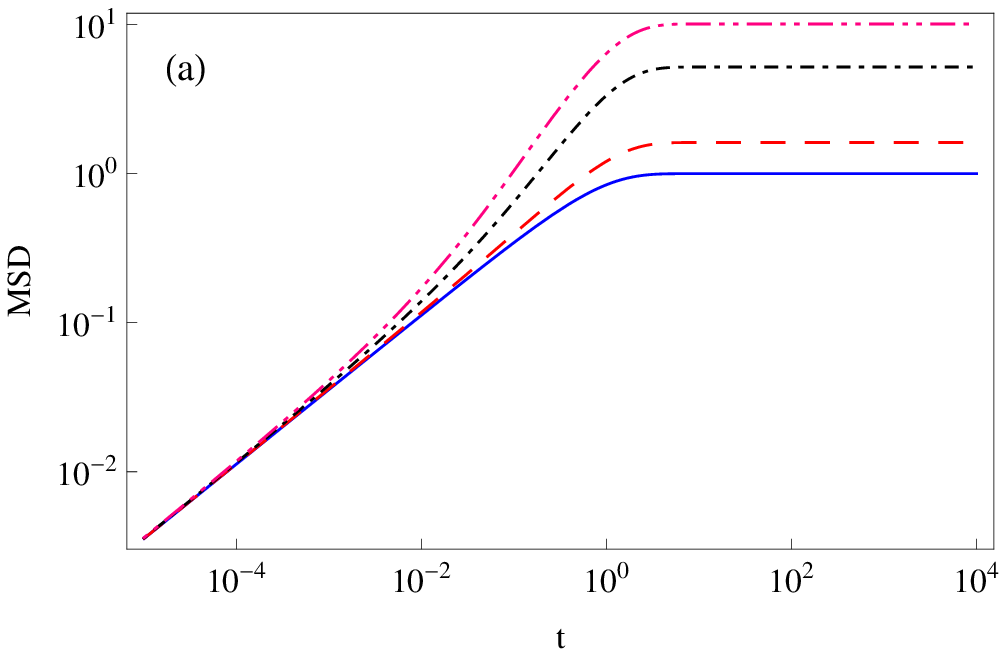}
\includegraphics[width=8cm]{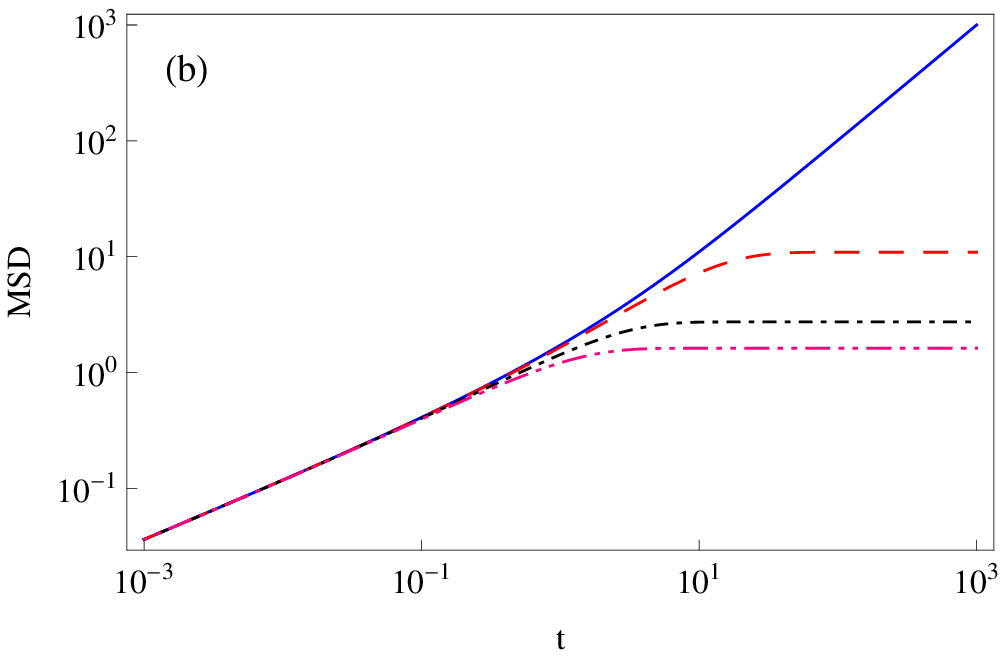}
\caption{MSD (\ref{msd final}) as function of time $t$ for (a) resetting rate $r=1$
and potential strength $U_0=0$ (blue solid line), $U_0=1$ (red dashed line), $U_0=5$
(black dot-dashed line), $U_0=10$ (violet dot-dot-dashed line); (b) for $U_0=1$ and
$r=0$ (blue solid line), $r=0.1$ (red dashed line), $r=0.5$ (black dot-dashed line),
$r=1$ (violet dot-dot-dashed line). We set $D_x=1$ and $D_y=1$.}
\label{fig2}
\end{figure}

From the final result (\ref{msd final}) for the MSD we observe a saturation in the
long time limit,
\begin{equation}
\langle x^2(t)\rangle\simeq\frac{D_x}{2D_y}U_0\frac{1+\Delta_{r}}{r},
\end{equation}
which occurs due to the resetting of the particle, while in the short time limit, we
observe the subdiffusive behaviour
\begin{equation}
\langle x^2(t)\rangle\sim2\frac{D_x}{\sqrt{D_y}}\frac{t^{1/2}}{\Gamma(1/2)},
\end{equation}
typical for free diffusion in a comb, since both resetting and potential do not
affect the particle dynamics at short times. A graphical representation of the
MSD is shown in figure~\ref{fig2}. In figure \ref{fig2}(a) we observe the transition
from subdiffusion, $\simeq t^{1/2}$, to the saturation plateau effected by resetting,
for different values of the potential energy $U_0$. Figure~\ref{fig2}(b) shows the
behaviour of the MSD for fixed potential strength, $U_0=1$, and different values of
the resetting rate $r$. For $r=0$ normal diffusion is observed in the long time
limit (blue solid line), which occurs due to the confining potential in the fingers. 

We finally write down the Fokker-Planck equation for the marginal PDF along the
branches, $p_{r,2}(y,t)=\int_{-\infty}^{\infty}p_r(x,y,t)\,dx$, in the form
\begin{equation}
\label{fpe y}
\frac{\partial}{\partial t}p_{r,2}(y,t)=\left(\frac{\partial}{\partial y}V'(y)
+D_y\frac{\partial^2}{\partial y^2}\right)p_{r,2}(y,t)-r\,p_{r,2}(y,t)+r\delta(y),
\end{equation}
which is the diffusion equation with resetting in presence of the confining potential
\cite{rk2020}.

\section{Crossover to the steady state}
\label{sec3}

We now analyse the crossover dynamics to the steady state. We rewrite the PDF
(\ref{p1 final laplace U0}) as follows
\begin{eqnarray}
\nonumber
\tilde{p}_{r,1}(x,s)&=&\frac{1}{2}\sqrt{\frac{2\sqrt{D_y}}{D_x}}\frac{s^{-1}(s+r)
\left(s+r+\frac{U_0^2}{4D_y}-\frac{U_0^2}{4D_y}\right)^{-1/2}}{\sqrt{\left(s+r+
\frac{U_0^2}{4D_y}\right)^{1/2}+\frac{U_0}{2\sqrt{D_y}}}}\\
\nonumber
&&\times\exp\left(-\sqrt{\frac{2\sqrt{D_y}}{D_x}}{\left[\left(s+r+\frac{U_0^2}{4
D_y}\right)^{1/2}-\frac{U_0}{2\sqrt{D_y}}\right]^{1/2}}|x-x_0|\right)\\
&\equiv&\tilde{p}_{0,1}(x,s+r+U_0^2/[4D_y])+s^{-1}r\tilde{p}_{0,1}(x,s+r+U_0^2/
[4D_y]),\nonumber\\
\label{p1 final laplace U0 2}
\end{eqnarray}
where we split the fraction $s^{-1}(s+r)$. Performing the inverse Laplace transform,
we obtain
\begin{eqnarray}
\label{pr1 renewal}
\fl p_{r,1}(x,t)=\exp\left(-\left[r+\frac{U_0^2}{4D_y}\right]t\right)p_{0,1}(x,t)+
\int_0^tr\exp\left(-\left[r+\frac{U_0^2}{4D_y}\right]t'\right)p_{0,1}(x,t')dt'.\nonumber\\
\end{eqnarray}
Here, $p_{0,1}(x,t)$ is given by
\begin{eqnarray}
\label{p0(x,t) final}
\fl p_{0,1}(x,t)=\frac{1}{2}\sqrt{\frac{2\sqrt{D_y}}{D_x}}\mathscr{L}^{-1}\left\{
\frac{\exp\left(-\sqrt{\frac{2\sqrt{D_y}}{D_x}}\left(s^{1/2}-\frac{U_0}{2\sqrt{
D_y}}\right)^{1/2}|x-x_0|\right)}{\left(s-\frac{U_0^2}{4D_y}\right)^{1/2}\left(
s^{1/2}+\frac{U_0}{2\sqrt{D_y}}\right)^{1/2}}\right\}.
\end{eqnarray}
This Laplace inversion of $p_{0,1}(x,t)$ for arbitrary $U_0\neq0$ is not
straightforward. However when $U_0=0$, this procedure is feasible. Therefore,
for the clarity of the analysis we first consider this simplified case in
absence of confinement in the branches. Then considering the simplified
asymptotic form of the PDF $p_{0,1}(x,t)$ we will be able to compare with
the difference in the presence of confinement, $U_0\neq 0$.

\subsection{The case confinement-free branches ($U_0=0$)}

In absence of the potential ($U_0=0$), the result of the Laplace
inversion in equation (\ref{p0(x,t) final}) is exact and expressed in the form
\begin{eqnarray}
\nonumber
p_{0,1}(x,t)&=&\frac{1}{2}\sqrt{\frac{2\sqrt{D_y}}{D_x}}\mathscr{L}^{-1}\left\{
s^{-3/4}\exp\left(-\sqrt{\frac{2\sqrt{D_y}}{D_x}}s^{1/4}|x-x_0|\right)\right\}\\
&=&\frac{1}{2}\sqrt{\frac{2\sqrt{D_y}}{D_x}}t^{-1/4}H_{1,1}^{1,0}\left[\left.
\sqrt{\frac{2\sqrt{D_y}}{D_x}}\frac{|x-x_0|}{t^{1/4}}\right|\begin{array}{l}
(3/4,1/4)\\(0,1)\end{array}\right],
\label{p01 final}
\end{eqnarray}
where $H_{p,q}^{m,n}(z)$ is the Fox $H$-function \cite{saxena_book}. Here we used
the identity
\begin{equation}
e^{-z}=H_{0,1}^{1,0}\left[z\left|\begin{array}{l}-\\(0,1)\end{array}\right.\right]
\end{equation}
and the inverse Laplace transform
\begin{eqnarray}
\mathscr{L}^{-1}\left\{s^{-\rho}H_{p,q}^{m,n}\left[a\,s^{\sigma}\left|\begin{array}{l}
(a_p,A_p)\\(b_q,B_q)\end{array}\right.\right]\right\}=t^{\rho-1}H_{p+1,q}^{m,n}\left[
\frac{a}{t^{\sigma}}\left|\begin{array}{l}(a_p,A_p),(\rho,\sigma)\\(b_q,B_q)
\end{array}\right.\right].\nonumber\\
\end{eqnarray}

This density form can be employed to evaluate the distribution $p_{r,1}(x,t)$
in the presence of resetting,
\begin{equation}
\label{renewal}
p_{r,1}(x,t)=e^{-rt}p_{0,1}(x,t)+\int^t_0dt're^{-rt'}p_{0,1}(x,t'),
\end{equation}
where the first term is given by equation (\ref{p01 final}) multiplied by $e^{-rt}$.

For further analysis it is convenient to use the asymptotic form for large argument
of the Fox $H$-function in equation (\ref{p01 final}). We find the non-Gaussian form
\cite{rangarajan}
\begin{equation}
\label{p01 final asympt}
p_{0,1}(x,t)\sim\exp\left(-\frac{3}{2^{8/3}}\left[\frac{a|x-x_0|}{t^{1/4}}\right]
^{4/3}\right).
\end{equation}
Substituting this expression into the integral in the renewal equation (\ref{renewal})
and focusing on the long time limit we have
\begin{equation}
\label{renewal integral ldf}
\int_0^te^{-rt'}p_{0,1}(x,t')dt'\approx\int^1_0d\tau\exp\left(-t\,\Phi(\tau,|x-x_0|/t)
\right),
\end{equation}
where 
\begin{eqnarray}
\label{Phi}
\Phi(\tau,|x-x_0|/t)=r\tau+\frac{3a^{4/3}}{2^{8/3}}\left(\frac{|x-x_0|}{t}\right)^{
4/3}\tau^{-1/3},\,\,\, a=\sqrt{2\sqrt{D_y}/D_x}.\nonumber\\
\end{eqnarray}
We evaluate the integral in the Laplace approximation \cite{arfken}, which requires evaluation
of the minimum of $\Phi$, defined as $0=\frac{d}{d\tau}\Phi|_{\tau=\tau_0}$,
such that $\tau_0=\frac{1}{4r^{3/4}}\frac{a|x-x_0|}{t}$. Physically, this corresponds to the relaxation behavior of $p_{r,1}(x; t)$ with the saddle point $\tau_0$ determining the spatial region in which relaxation has been achieved, at time t. Outside the region the
system is still in a transient state, and corresponds to the saddle point lying outside the unit interval. Thus, in the
transient space-time region, the maximal contribution to the integral comes from the end point at  $\tau = 1$. Therefore, within this Laplace
approximation, the large deviation form for the PDF $p_{r,1}(x,t)$ can be written as follows
\begin{equation}
p_{r,1}(x,t)\sim\exp\left(-t\,I_r\left(\frac{|x-x_0|}{t}\right)\right),
\end{equation}
where the large deviation function is
\begin{equation}
I_r\left(\frac{|x-x_0|}{t}\right)=\left\{\begin{array}{ll}ar^{1/4}\frac{|x-x_0|}{t},
& |x-x_0|<\frac{4r^{3/4}}{a}t, \\[0.2cm]
r+\frac{3a^{4/3}}{2^{8/3}}\left(\frac{|x-x_0|}{t}\right)^{4/3}, & |x-x_0|>\frac{4r
^{3/4}}{a}t.\end{array}\right.
\label{ldf_ir}
\end{equation}

\begin{figure}
\includegraphics[width=8cm]{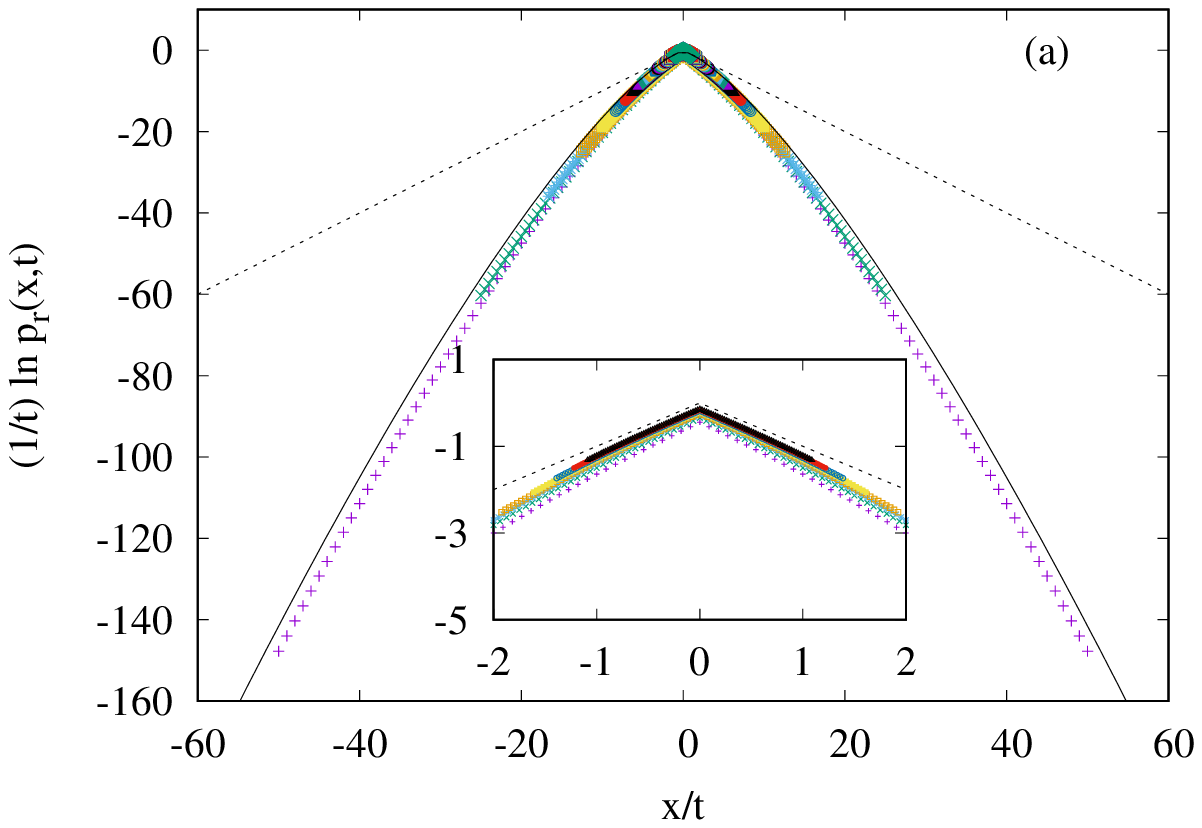}
\includegraphics[width=8cm]{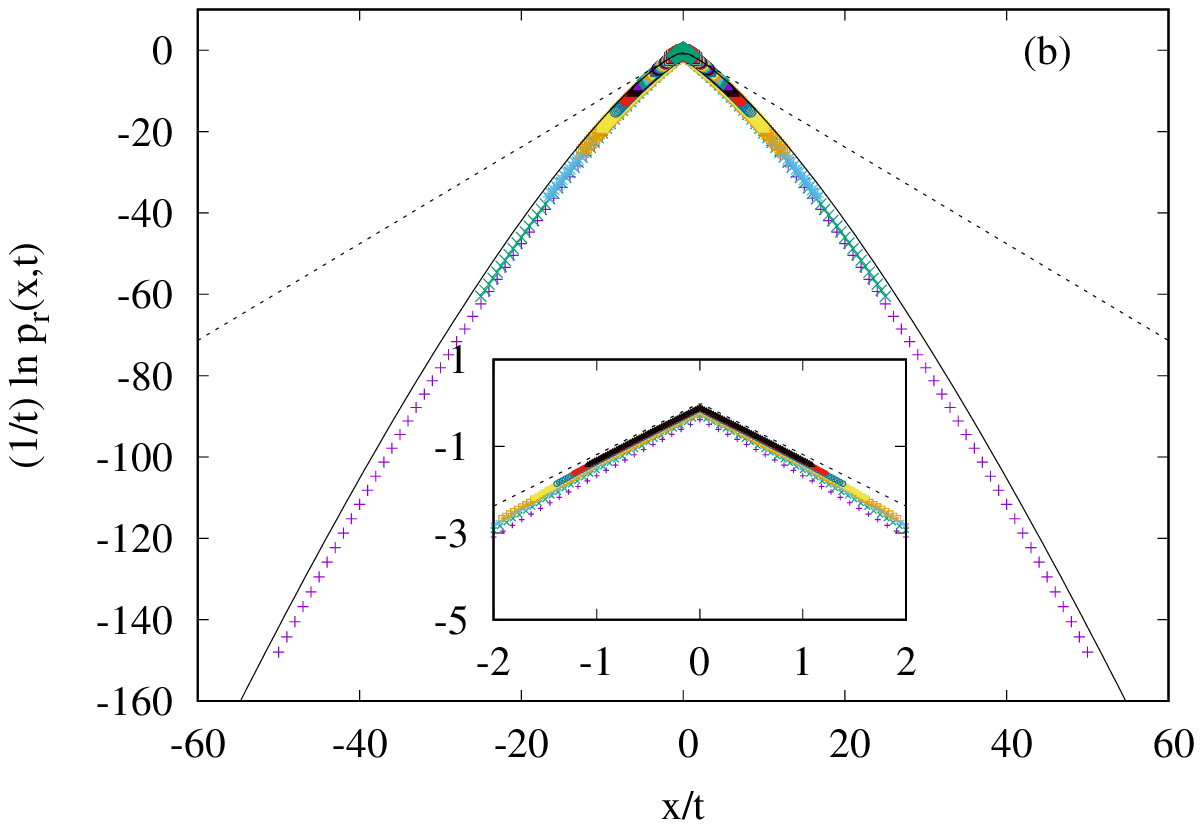}\\
\includegraphics[width=8cm]{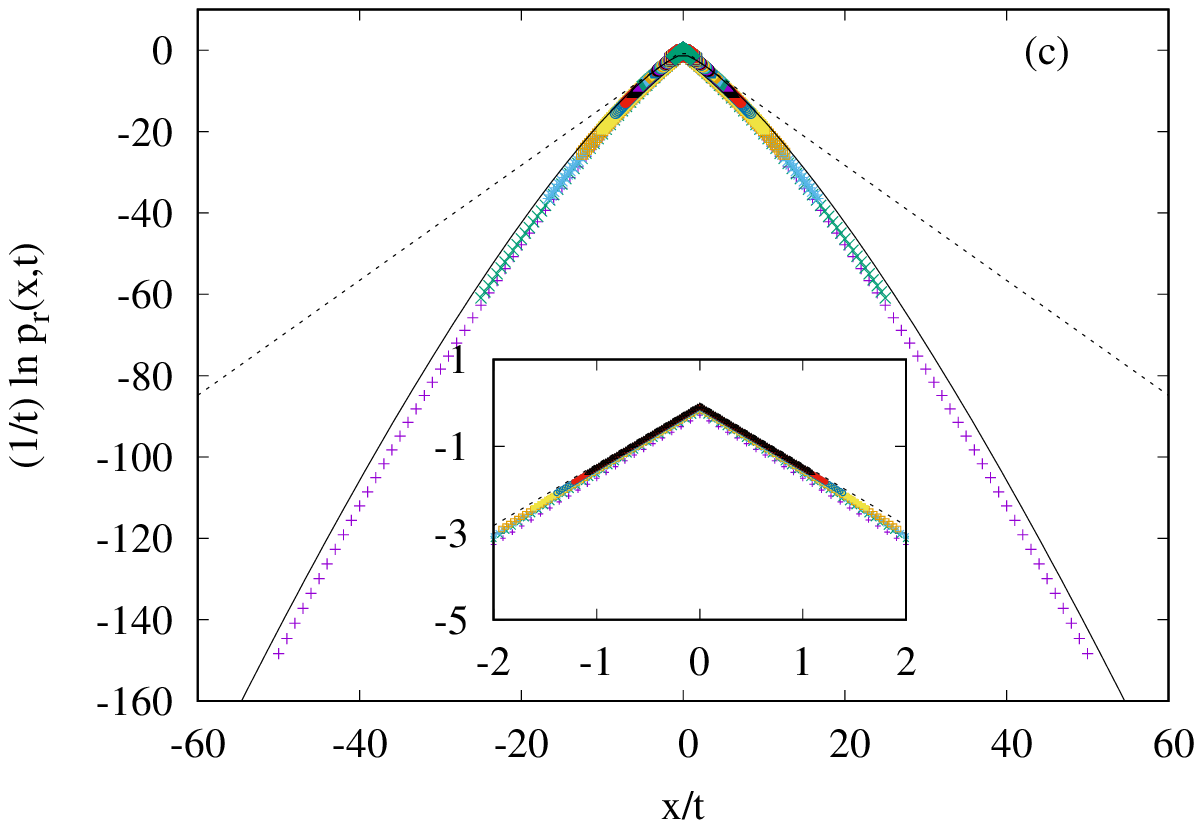}
\includegraphics[width=8cm]{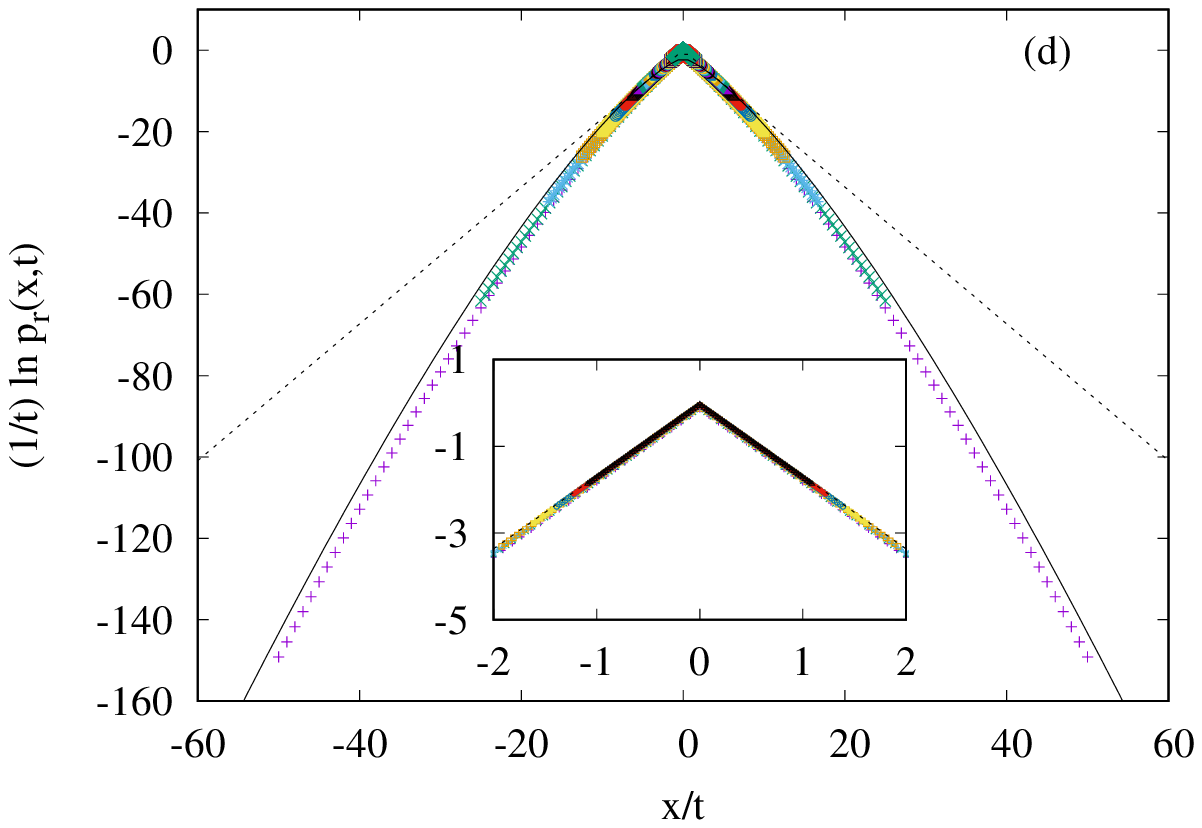}
\caption{Logarithm of the PDF scaled by time, $(1/t)\log{p_{r,1}(x,t)}$,
vs $x/t$ for (a) $r=1/4$, (b) $r = 1/2$, (c) $r = 1$, and (d) $r = 2$. The
data are obtained by numerical inverse Laplace transform of the PDF (\ref{p1
final laplace U0 2}) in Mathematica for $U_0=0$.  The inset in the figures
shows blow-up for small values of argument $|x|/t$. The black dotted line
and the black solid lines indicate the two forms for the large deviation
function $I_r$, see Eq.~(\ref{ldf_ir}).
}
\label{fig_ldf}
\end{figure}

From the form of the large deviation function (\ref{ldf_ir}) it is evident that there occurs a qualitative change in the density profile $p_{r,1}(x,t)$ at a space-time point defined by $\tau_0 < 1$. This demarcates a ``light-cone'' region within which relaxation has been achieved and outside it the system is still relaxing. This relaxation
behavior is, however, slower than the case of a Brownian motion
relaxing to its nonequilibrium steady state under resetting \cite{schehr}.
The reason for this difference is that unlike Brownian motion on a line, a random walk on a two dimensional comb is subdiffusive ($U_0 = 0$). And hence,
even though resetting is the common mechanism responsible for bringing about
relaxation in both cases, the rate of relaxation, which is governed primarily
by systemic details, is significantly different. Here we note that even though the stationary distribution in case of a diffusion in combs with resetting has been analysed before \cite{prr}, this is the first time to explicitly find the corresponding large deviation function.

In order to verify our analytical estimates of the large deviation approximation of $p_r(x,t)$, we numerically invert the Laplace transform $\tilde{p}_r(x,s)$ for different values of the resetting rate $r$ as presented in figure~\ref{fig_ldf}. It is evident from the graphs that the numerical estimates very nicely corroborate our analytical results.

\subsection{Presence of confinement in the branches ($U_0>0$)}

\begin{figure}
\centering{\includegraphics[width=8cm]{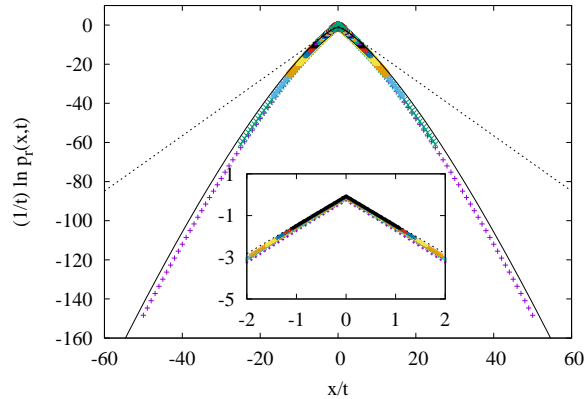}}
\caption{Logarithm of the PDF scaled by time, $(1/t)\log{p_{r,1}(x,t)}$, vs $x/t$, for $r=1$ and $U_0=1$. The black solid line and the black dotted line
are drawn following a scaling form similar to the large deviation function
in (\ref{ldf_ir}). It is to be noted, however, that these are numerically
obtained results via analogy from the case $U_0 = 0$, see text.
}
\label{ldf_u0eq1}
\end{figure}

In the presence of the confining potential, $U_0>0$, one cannot perform an analytical
Laplace inversion of the PDF (\ref{p1 final laplace U0 2}). We therefore resort to
numerical Laplace inversion of expression (\ref{p1 final laplace U0 2}), as shown in
figure~\ref{ldf_u0eq1}.

In order to understand the result in figure~\ref{ldf_u0eq1}
let us compare equations (\ref{pr1 renewal}) and (\ref{renewal}), which
respectively are renewal equations for motion under resetting on a comb with
and without confining branches. A careful inspection of the two equations
makes it immediately evident that the confinement $U_0$ tends to modify the
resetting rate $r$, except for the common prefactor of the integrals in
(\ref{pr1 renewal}) and (\ref{renewal}). This is because both resetting and
confinement have the effect of bringing the particle towards the backbone
with one minor difference. Whereas resetting is instantaneous and takes the
particle from anywhere on the comb to its initial location, the effect of
confinement is non-instantaneous. The Brownian particle spends some
time in its excursion along the branches before returning to the backbone.
Furthermore, the location of return along the backbone due to confinement
is not necessarily its initial location. Notwithstanding these slight differences,
we see in figure~\ref{ldf_u0eq1} that the scaling function rendering
the collapse of $p_{r,1}(x,t)$ at different times exhibits a behavior similar
to case of nonconfining branches.

\section{First-passage times along the backbone}
\label{sec4}

We now turn to consider the first-time passage statistic along the backbone, by
placing an absorbing boundary at $x=L>0$, i.e., $p_1(L,t)=0$. Without loss of
generality we choose $x_0<L$. The equation of motion for the density function
$p_1(x,t)$ in Laplace space along the backbone in absence of resetting follows
from equation (\ref{p1 eq laplace}),
\begin{equation}
s\tilde{p}_1(x,s)-\delta(x-x_0)=\frac{U_0D_x}{4D_y}(1+\Delta_s)\frac{\partial^2
\tilde{p}_1}{\partial x^2},
\end{equation}
to be augmented with the boundary condition $p_1(L,t)=0$. We rephrase this
expression as
\begin{equation}
\frac{\partial^2\tilde{p}_1}{\partial x^2}-As\tilde{p}_1=-A\delta(x-x_0),
\end{equation}
where $A=4D_y/[U_0D_x(1+\Delta_s)]$. Now, the auxiliary equation for the
case $x \neq x_0$ for the above
differential equation is $0=m^2-As$, implying $m=\pm\sqrt{As}$. We thus obtain
\begin{equation}
\tilde{p}_1(x,s)=\left\{\begin{array}{ll}
b_+\exp\left(\sqrt{As}x\right)+b_-\exp\left(-\sqrt{As}x\right), & x<x_0, \\[0.2cm]
c_+\exp\left(\sqrt{As}x\right)+c_-\exp\left(-\sqrt{As}x\right), & x>x_0.
\end{array}\right.
\end{equation}
Since $-\infty<x\leq L$ the requirement for physically meaningful solutions in the
region $x<x_0$ is $b_-=0$. Continuity of the solution at $x=x_0$ and discontinuity
of the derivative owing to the probability source at $x=x_0$ provide us with two
relations between the parameters $b_+$ and $c_{\pm}$,
\begin{eqnarray}
\nonumber
c_+\exp\left(\sqrt{As}x_0\right)+c_-\exp\left(-\sqrt{As}x_0\right)-b_+\exp\left(
\sqrt{As}x_0\right)&=&0,\\
c_+\exp\left(\sqrt{As}x_0\right)-c_-\exp\left(-\sqrt{As}x_0\right)-b_+\exp\left(
\sqrt{As}x_0\right)&=&-\sqrt{A/s}.\nonumber\\
\end{eqnarray}
In order to determine the value of these constants in terms of the system parameters
we need one more relation, provided by the absorbing boundary condition at $x=L>x_0$,
i.e., $\tilde{p}_1(L,s)=0$. Along with the previous two relations, this constraint
fixes the parameters uniquely, and we obtain the density
\begin{equation}
\tilde{p}_1(x,s)=\left\{
\begin{array}{ll}
\displaystyle\sqrt{\frac{A}{s}}\sinh\left[\sqrt{As}(L-x_0)\right]\exp\left(\sqrt{
As}(x-L)\right), & x<x_0, \\[0.4cm]
\displaystyle
\sqrt{\frac{A}{s}}\sinh\left[\sqrt{As}(L-x)\right]\exp\left(\sqrt{As}(x_0-L)\right), & x>x_0.\end{array}\right.
\end{equation}
Note that in the presence of the absorbing boundary the quantity $p_1(x,t)$ is no
longer a PDF, as the cumulative (survival) probability becomes a decaying function
of time. We then are in the position to derive the first-passage time density (FPTD)
\begin{equation}
\wp_1(t)=-\frac{d}{dt}\int_{-\infty}^Lp_1(x,t)dx,
\end{equation}
where the integral on the right hand side represents the survival probability.
In Laplace domain,
\begin{eqnarray}
\tilde{\wp}_1(s)&=&-\int^L_{-\infty}\Big[s\tilde{p}_1(x,s)-\delta(x-x_0)\Big]dx=\left.
-\frac{1}{A}\frac{\partial\tilde{p}_1}{\partial x}\right|_{x=L}\nonumber\\&=&\exp\left((x_0-L)
\sqrt{\frac{4D_ys}{U_0D_x(1+\Delta_s)}}\right),
\end{eqnarray}
where $\Delta_s=\sqrt{1+\frac{4sD_y}{U^2_0}}$. After Laplace inversion, the first-passage time reads
\begin{eqnarray}
\nonumber
\wp_1(t)&&=\mathscr{L}^{-1}\left\{\exp\left(-\sqrt{\frac{4D_ys(\Delta_s-1)}
{U_0D_x(\Delta_s^2-1)}}(L-x_0)\right)\right\}\nonumber\\&&=\mathscr{L}^{-1}\left\{\exp\left(
-\sqrt{\frac{U_0(\Delta_s-1)}{D_x}}(L-x_0)\right)\right\}\nonumber\\
&&=\mathscr{L}^{-1}\left\{\exp\left(-\sqrt{\frac{2\sqrt{D_y}}{D_x}}\left[\left(
s+\frac{U_0^2}{4D_y}\right)^{1/2}-\frac{U_0}{2\sqrt{D_y}}\right]^{1/2}(L-x_0)
\right)\right\}.\nonumber\\
\label{F1}
\end{eqnarray}

\begin{figure}
\includegraphics[width=8cm]{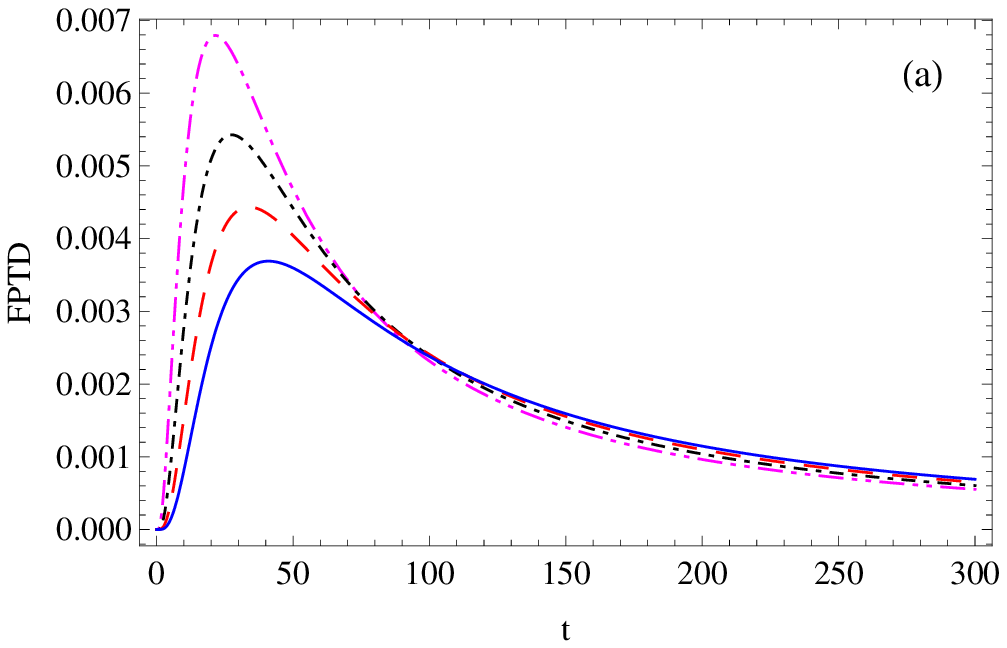}
\includegraphics[width=8cm]{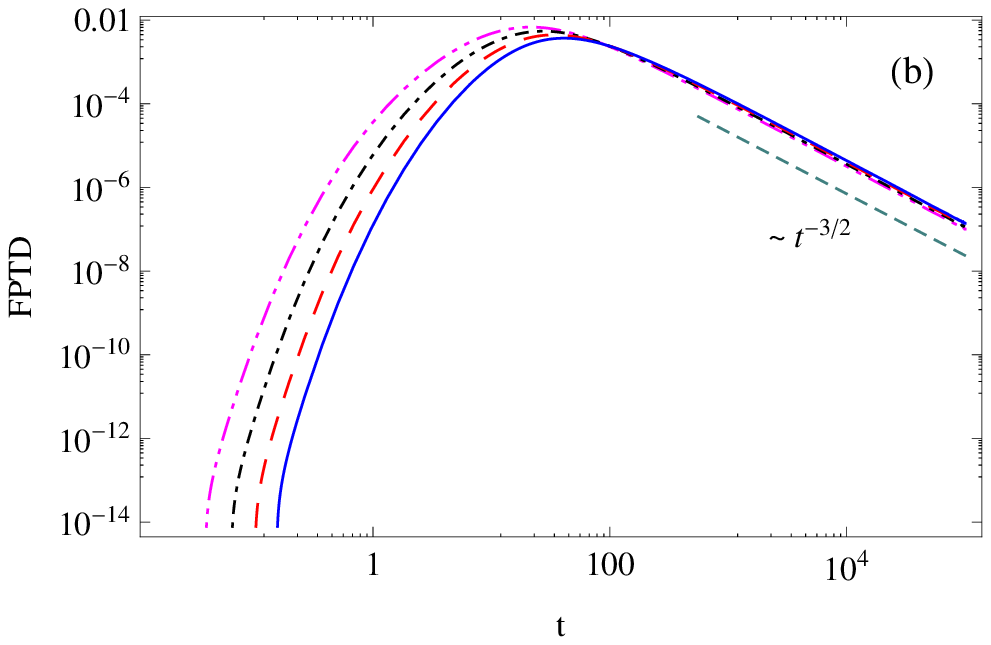}
\caption{FPTD (\ref{F1}) as function of time $t$ for $D_x=1$, $D_y=1$, $U_0=1$, $L=10$ and $x_0=-1$ (blue solid line), $x_0=0$ (red dashed line), $x_0=1$ (black dot-dashed line) and $x_0=2$ (violet dot-dot-dashed line) (a) Linear-linear plot, (b) Log-log plot.}
\label{fig_FPTD}
\end{figure}

For the long time limit, we find
\begin{eqnarray}
\wp_{1}(t)&&\stackrel[t\rightarrow\infty]{}{\sim}\left.\mathscr{L}^{-1}\left\{\exp\left(-\sqrt{\frac{2D_y}{D_xU_0}}(L-x_0)s^{1/2}\right)\right\}\right|_{s\rightarrow0}\nonumber\\&&\sim\mathscr{L}^{-1}\left\{1-\sqrt{\frac{2D_y}{D_xU_0}}(L-x_0)s^{1/2}\right\}. 
\end{eqnarray}
Therefore, 
\begin{eqnarray}
s^{-1}\tilde{\wp}_{1}(s)&\stackrel[s\to0]{}{\sim} s^{-1}-\sqrt{\frac{2D_y}{D_xU_0}}(L-x_0)s^{-1/2},
\end{eqnarray}
from where, by inverse Laplace transform, it follows that
\begin{eqnarray}
\fl\int_{0}^{t}\wp_{1}(t')dt'\sim1-\sqrt{\frac{2D_y}{D_xU_0}}(L-x_0)\frac{t^{-1/2}}{\Gamma(1/2)} \quad \rightarrow \quad \wp_{1}(t)\stackrel[t\rightarrow\infty]{}{\sim}\sqrt{\frac{D_y}{2D_xU_0}}(L-x_0)\frac{t^{-3/2}}{\Gamma(1/2)}.\nonumber\\
\end{eqnarray}
A graphical representation of the FPTD is given in fiigure~\ref{fig_FPTD}. It is evident that in the long time limit the FPTD behaves as $\wp_{1}(t)\simeq t^{-3/2}$


\begin{figure}
\includegraphics[width=8cm]{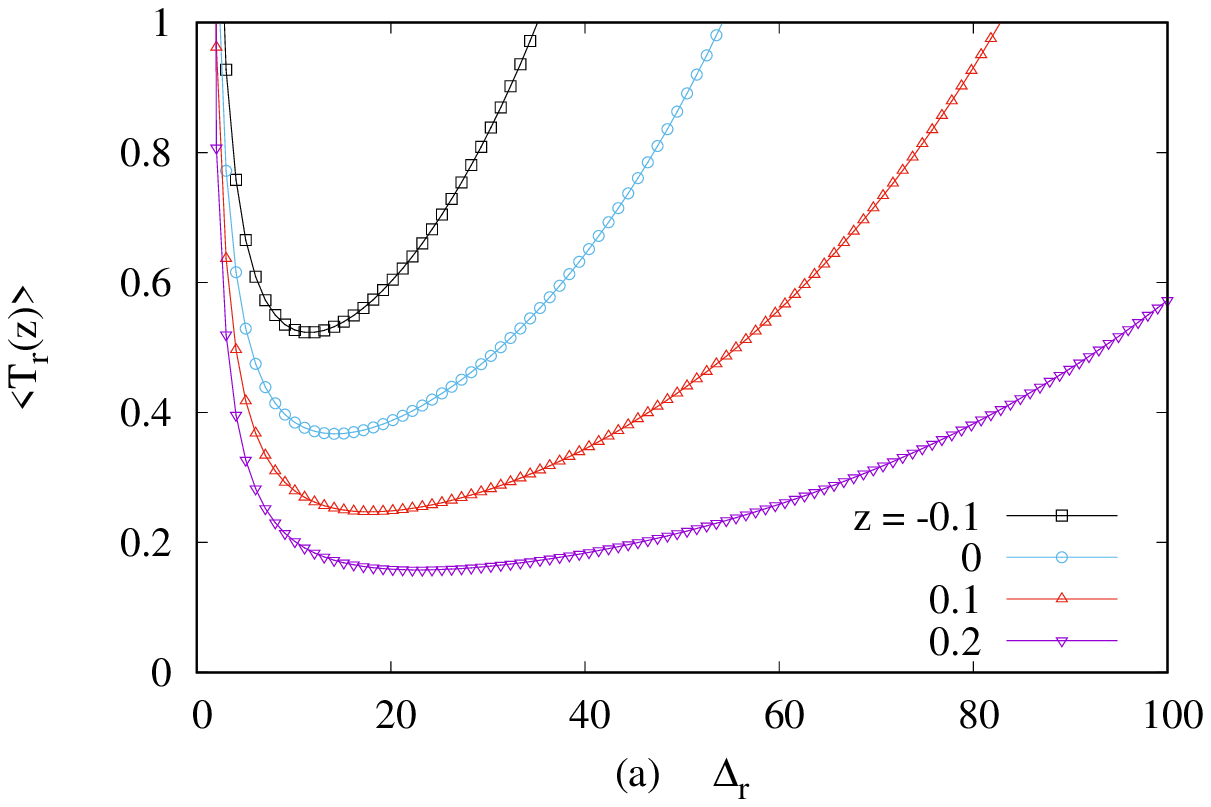}
\includegraphics[width=8cm]{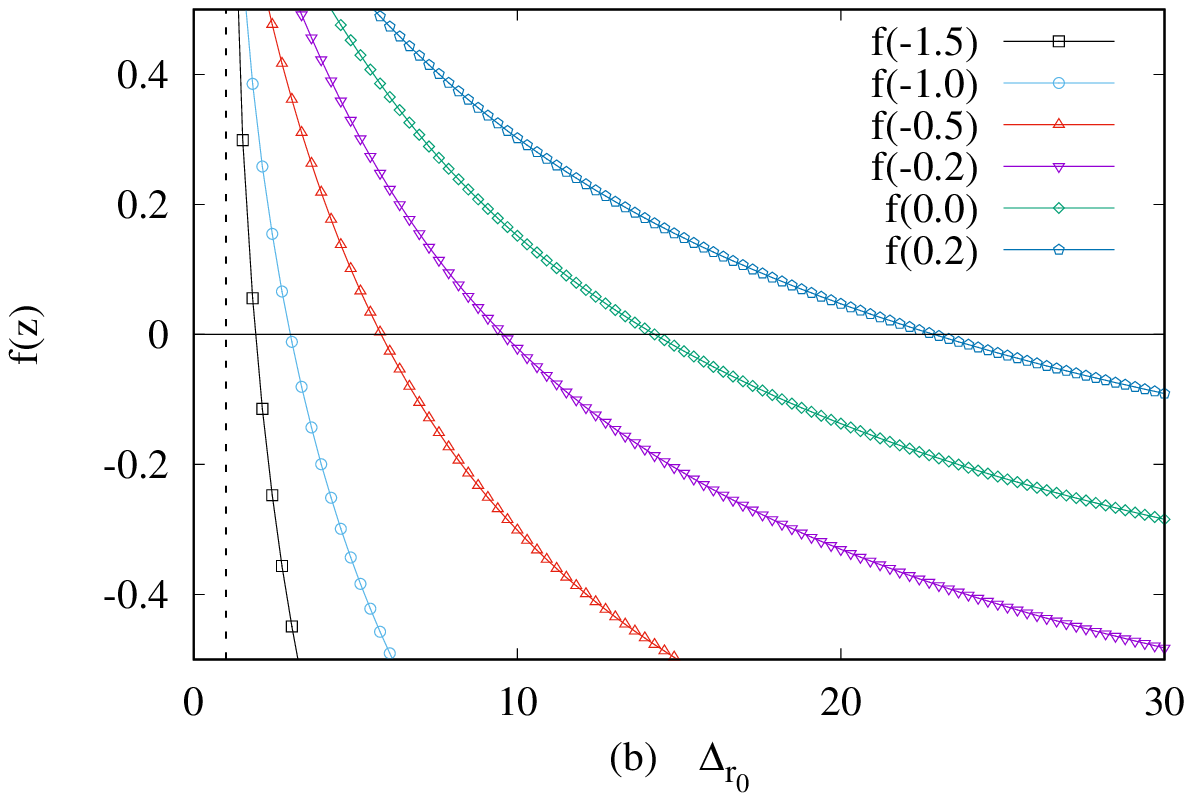}
\caption{(a) Non-monotonic dependence of the mean first-passage time on the resetting
rate $r$ and (b) zero-crossings of $f(z,x)=0$. The vertical dashed line in panel
(b) indicates $\Delta_r=1$, signifying the fact that as the (normalised) resetting
location $z=x_0/L$ approaches large negative values the optimal resetting rate $r_0$
approaches zero.
}
\label{fpt}
\end{figure}

The mean first-passage time for normal diffusion on a semi-infinite line is
infinite \cite{redner}. The same divergence will therefore occur in our comb
structure for the motion along the semi-infinite domain on the backbone in
absence of resetting. In that case we either have a crossover from subdiffusion
to normal diffusion when the diffusion in the branches is confined ($U_0>0$), or
continuing subdiffusion when there is no confinement, see also \cite{evans,resto}.
Once we switch on the resetting dynamics, however, we expect the mean first-passage
time to be finite. Using the results of \cite{reuveni} we find that expression
(\ref{F1}) for the first-passage time density in absence of resetting helps us
evaluate the mean first-passage time when resetting occurrs,
\begin{equation}
\label{Tr}
\langle T_r(x_0)\rangle=\frac{1}{r}\left[\exp\left((L-x_0)\sqrt{\frac{U_0}{D_x}
(\Delta_r-1)}\right)-1\right],
\end{equation}
where $\Delta_r=\sqrt{1+\frac{4rD_y}{U^2_0}}$. The divergence of the mean first-passage time in absence of resetting from this expression is obvious when we take the
limit $r\to0$. We also note the rapid growth of the mean first-passage time when the
particle is rapidly reset to its initial location. In such a case the particle has
an increasingly smaller chance to ever reach the absorbing boundary before the next
reset. According to expression (\ref{Tr}) the divergence of $\langle T_r(x_0)\rangle$
corresponds to a pole of the form $1/r$ whereas the divergence for large $r$ is
exponential. The non-monotonic behaviour of the mean first-passage time with the
resetting rate is shown in figure~\ref{fpt} where we plot $\langle T_r(x_0)\rangle$
as function of $\Delta_r$ for different (normalised) resetting locations $z=x_0/L$.

Apart from these immediate conclusions it is interesting to look at the
behaviour of the mean first-passage time $\langle T_r(x_0)\rangle$ as a
function of the resetting rate $r$ in more detail. To this end we introduce
two dimensionless quantities, $\tau_0 =2D_y/U^2_0$ and $\mu=U_0L/(D_x/L)$
representing, respectively, a dimensionless time-scale and the ratio of the
energy barrier to diffusion strength. Now, without any loss of generality
we can choose the dimensionless time-scale as unity, i.e., $\tau_0=1$. In
addition, as the orientations of the confining potential and the backbone
are orthogonal to each other, we are at liberty to independently choose
the values of $U_0$ and $D_x$. For simplicity we therefore choose $\mu=1$,
without limiting generality. Then the mean first-passage time simplifies to
\begin{equation}
\label{trnorm}
\langle T_r(z)\rangle=\frac{2}{\Delta^2_r-1}\left[\exp\left((1-z)\sqrt{\Delta_r-1}
\right)-1\right],
\end{equation}
where now $\Delta_r=\sqrt{1+2r}$. It is evident from this expression that the
mean first-passage time to the absorbing wall in the presence of resetting exists for
every $z\le1$ with $\langle T_r(1)\rangle =0$. The latter result is obvious,
as the initial position coincides with the absorbing boundary. Expression
(\ref{trnorm}) also allows the calculation of the optimal resetting rate $r_0$,
at which the mean first-passage time is minimal, $\frac{d}{dr}\langle T_r(x_0)
\rangle|_{r = r_0}=0$, resulting in the transcendental equation
\begin{eqnarray}
f(z)\equiv\frac{4\Delta_{r_0}}{(\Delta_{r_0}+1)\sqrt{\Delta_{r_0}-1}}\left[1-\exp
\left(-(1-z)\sqrt{\Delta_{r_0}-1}\right)\right]-(1-z)=0,\nonumber\\
\end{eqnarray}
which uniquely fixes $\Delta_{r_0}$ for a given value of $z$. Numerical analysis
of this relation between the optimal $\Delta_{r_0}$ and function $f(z)$ for the
corresponding resetting position $z$ as shown in figure~\ref{fpt} demonstrates
that the optimal resetting rate $r_0$ approaches zero as the reset location $z$
takes large negative values. In other words, the optimal resetting rate exhibits
a vanishing transition given that the mean first-passage time in absence of
resetting is infinite.

\section{Conclusions}
\label{sec5}

SR is a phenomenon with almost ubiquitous relevance in a large range of systems,
from diffusion controlled regulation in molecular biological processes to the
search of higher animals for food. We here combined SR with the well established
comb structure, a widely used model for loopless heterogeneous structures, with applications
ranging from biologically relevant cases such as nerve fibres or blood vessels to
aquifer backbones in groundwater dispersion. In our two-dimensional comb model we
applied a confining potential of strength $U_0$, mimicking a finite length of the
comb's branches such that the mean residence time in these branches is kept finite.
On top of the diffusivities $D_x$ and $D_y$ along the comb's backbone and the
branches, respectively, our system is therefore described by two additional
relevant parameters, the confinement strength $U_0$ and the resetting rate $r$.

We demonstrated that while in absence of resetting a crossover occurs from
initial subdiffusion to long time normal diffusion with Gaussian PDF (for the
normal diffusion in absence of resetting see also an alternative viewpoint of
the backbone diffusion in \ref{app2}), in the presence of resetting
the initial subdiffusion eventually crosses over to a non-equilibrium steady
state behaviour characterised by a plateau of the MSD. Depending on the choice
of parameters, an intermediate normal diffusion regime may be observed. The PDF
in the non-equilibrium steady state was shown to be of stretched exponential
shape. We analysed the crossover dynamics to the steady state based on the large
deviation function (\ref{ldf_ir}) using the asymptotic Laplace approximation
method. This result also shows that the space-time region is now
demarcated by a light-cone within which the system has relaxed to its
nonequilibrium steady state, similar to the case of Brownian motion
on a line. Outside this light-cone region, however, the rate of relaxation
is slower in comparison to that of Brownian motion under resetting.
This is because in the present geometry the particle tends to spend
a finite amout of time along the branches rendering the relaxation,
which is governed by the systemic details, to be achieved at a slower
pace. 

We also investigated the first-passage dynamics along the backbone. In particular
we investigated the first-passage behaviour as function of the resetting rate and
the amplitude of the confining potential. We calculated the mean first-passage
time and the optimal resetting rate, at which the minimal first-passage time is
obtained. Good agreement with a numerical evaluation is observed.

\ack{RM \& TS acknowledges financial support by the German Science Foundation (DFG,
Grant number ME 1535/12-1). RM also thanks the Foundation for Polish Science
(Fundacja na rzecz Nauki Polskiej) for support within an Alexander von Humboldt
Polish Honorary Research Scholarship. TS was supported by the Alexander von
Humboldt Foundation (Grant No. MKD 1205769 GF-E). TS also acknowledges support from the bilateral Macedonian-Chinese research project 20-6333, funded under the intergovernmental Macedonian-Chinese agreement.}

\appendix

\section{Coupled Langevin equation approach and subordination}
\label{app1}

From equation (\ref{eq general fl}) we find that the backbone's marginal PDF
satisfies
\begin{equation}
\label{eq general fl2}
\tilde{p}_1(k,s)=\frac{\frac{1}{s\tilde{\eta}(s)}}{\frac{1}{\tilde{\eta}(s)}
+\mathcal{D}k^2},
\end{equation}
where $\mathcal{D}=\frac{D_x}{2\sqrt{D_y}}$. Alternatively, in integral form,
\begin{eqnarray}
\tilde{p}_1(k,s)&=&\frac{1}{s\tilde{\eta}(s)}\int_0^{\infty}\exp\left(-u(1/\tilde{
\eta}(s)+\mathcal{D}k^2)\right)du\nonumber\\&=&\int_0^{\infty}e^{-u\mathcal{D}k^2}\tilde{h}(u,
s)du,
\end{eqnarray}
where
\begin{equation}
\tilde{h}(u,s)=\frac{1}{s\tilde{\eta}(s)}e^{-u/\tilde{\eta}(s)}.
\end{equation}
From inverse Fourier-Laplace transform we find \cite{csf}
\begin{equation}
p_1(x,t)=\int_0^{\infty}\frac{\exp(-x^2/[4\mathcal{D}u])}{\sqrt{4\pi\mathcal{D}u}}
h(u,t)\,du.
\end{equation}
The function $h(u,t)$ is called the subordinator\footnote{Note that $h(u,t)$ is
normalised since
\[
\int_0^{\infty}h(u,t)du=\mathscr{L}^{-1}\left\{\int_0^{\infty}\tilde{h}(u,s)du\right\}=
\mathscr{L}^{-1}\left\{\int_0^{\infty}\frac{e^{-u/\tilde{\eta}(s)}}{s\tilde{\eta}(s)}du\right\}
=\mathscr{L}^{-1}\left\{\frac{1}{s}\right\}=1.
\]
}
which re-expresses the random process governed by the generalised diffusion
equation (\ref{p1 eq}) in physical time $t$ to the Wiener process with
Gaussian PDF $f(x,u)=(4\pi\mathcal{D}u)^{-1/2}\exp\left(-x^2/[4\mathcal{D}u]
\right)$, in terms of the operational time $u$.

This result can in fact be obtained from CTRW theory by considering the stochastic
equations \cite{fogedby1}
\begin{eqnarray}
\left\lbrace\begin{array}{ll}
    \frac{d}{du}x(u)=\xi(u),  \\ \\
    \frac{d}{du}T(u)=\zeta(u), 
\end{array}
\right.
\end{eqnarray}
where $\xi(u)$ is a white Gaussian noise with zero mean and autocorrelation $\langle
\xi(u)\xi(u')\rangle=2\delta(u-u')$ while $\zeta(u)$ is a completely one-sided
L\'evy stable noise. This means that the random walk $x(t)$ is parametrised in terms
of the "number of steps" $u$. The inverse process $S(t)$ of the L\'evy process
$T(u)$ with characteristic function $\langle\exp(-sT(u))\rangle=\exp(-\Psi(s)u)$
represents a collection of first-passage times, $S(t)=\inf\{u>0:T(u)>t\}$
\cite{fogedby1}. Then the CTRW can be defined by the subordinated process $X(t)=
x(S(t))$. The PDF $h(u,t)$ of the inverse process $S(t)$ can be found from the
relation \cite{fogedby1}
\begin{equation}
h(u,t)=-\frac{\partial}{\partial u}\langle\Theta(t-T(u))\rangle,
\end{equation}
where $\Theta(z)$ is the Heaviside step function. Laplace transform then yields
\begin{eqnarray}
\tilde{h}(u,s)&=&-\frac{\partial}{\partial u}\frac{1}{s}\left<\int_0^{\infty}\delta(
t-T(u))e^{-st}dt\right>\nonumber\\&=&-\frac{\partial}{\partial u}\frac{1}{s}\langle e^{-sT(u)}
\rangle=-\frac{\partial}{\partial u}\frac{1}{s}e^{-\Psi(s)u}=\frac{\Psi(s)}{s}e^{
-\Psi(s)u}.
\end{eqnarray}
Therefore, 
\begin{equation}
p_1(x,t)=\langle\delta(x-X(t))\rangle=\langle\delta(x-X(S(t))\rangle=\int_0^{
\infty}f(x,u)h(u,t)dt,
\end{equation}
from where one can easily arrive at the generalised diffusion equation (\ref{p1 eq}),
when $\tilde{\Psi}(s)=1/\tilde{\eta}(s)$, where $\tilde{\eta}(s)$ is given by 
equation (\ref{eta comb potential}). The corresponding CTRW model represents a
random process with Gaussian jump length PDF and waiting time PDF in the Laplace
domain of the form $\tilde{\psi}(s)=(1+1/\tilde{\eta}(s))^{-1}\sim1-1/\tilde{\eta}
(s)$.

\section{Confinement along the branches}
\label{app2}

As a result of confinement along the branches, the particle tends to exhibit a
normal diffusive transport along the backbone at longer times. Furthermore the
potential along the $y$-axis branches results in a steady-state
\begin{equation}
p_{0,2,st}(y)\simeq\exp\left(-\frac{U_0}{D_y}|y|\right).
\end{equation}
If we look at distribution of the maxima of excursions along the $y$-branch, then
\begin{equation}
\fl P(M_n\ge y)=P(Y_1\ge y,\ldots,Y_n\ge y)=[P(Y\ge y)]^n\sim\exp\left(-\left[\frac{
nU_0}{D}\right]y\right),
\end{equation}
which implies that the maximal excursions along the confining branches are
exponentially distributed. In other words, the confinement effectively
confined diffusion in branch regions of finite length \cite{WhBa84}.

\end{document}